\documentclass{article}
\pdfoutput=1

\usepackage{amssymb,amsfonts,amsmath}
\usepackage{cite,enumerate,float}
\usepackage{color}
\usepackage{tikz}
\usetikzlibrary{arrows,snakes,backgrounds}

\def\be{\begin{eqnarray}}
\def\ee{\end{eqnarray}}
\def\nn{\nonumber}

\def\p{\partial}

\def\Tr{{\rm Tr}\,}

\definecolor{red}{rgb}{1,0,0}
\definecolor{orange}{rgb}{1,0.5,0}
\definecolor{violet}{rgb}{0.7,0,1}

\hoffset= - 1.0in         

\begin{document}

\begin{center}
\begin{small}
\hfill MIPT/TH-06/22\\
\hfill FIAN/TD-03/22\\
\hfill ITEP/TH-06/22\\
\hfill IITP/TH-05/22\\

\end{small}
\end{center}

\vspace{.5cm}

\begin{center}
\begin{Large}\fontfamily{cmss}
\fontsize{15pt}{27pt}
\selectfont
	\textbf{On combinatorial generalization(s) of Borel transform:
Averaging method in combinatorics of symmetric polynomials}
	\end{Large}
	
\bigskip \bigskip

\begin{large}
A. Mironov$^{b,c,d}$\,\footnote{mironov@lpi.ru; mironov@itep.ru},
A. Morozov$^{a,c,d}$\,\footnote{morozov@itep.ru}
\end{large}

\bigskip

\begin{small}
$^a$ {\it MIPT, Dolgoprudny, 141701, Russia}\\
$^b$ {\it Lebedev Physics Institute, Moscow 119991, Russia}\\
$^c$ {\it Institute for Information Transmission Problems, Moscow 127994, Russia}\\
$^d$ {\it NRC ``Kurchatov Institute'' - ITEP, Moscow 117218, Russia}\\
\end{small}
 \end{center}

\bigskip

\begin{abstract}
We elaborate on the recent suggestion to consider averaging of Cauchy identities for the
Schur functions over power sum variables.
This procedure has apparent parallels with the Borel transform,
only it changes the number of combinatorial factors like $d_R$ in the sums over Young diagrams
instead of just factorials in ordinary sums over numbers.
It provides a universal view on a number of previously known, but seemingly random identities.
\end{abstract}

\bigskip

\section{Introduction}

Borel transform \cite{Borel} is an important tool in the study of non-perturbative phenomena.
It is used to extract information from  divergent series
which typically arise in perturbative expansions.
Non-perturbative partition functions depend on extra parameters,
like the choice of integration contour, this dependence seems
to disappear in perturbative expansion, but is in fact traded for
divergence of the series.
New parameters appear as an ambiguity in the Borel transform.
All this is well seen already in the simplest example of the exponential Borel transform
\be
\sum_r   r!\cdot q^r  =
 \int_0^\infty \sum_r (qx)^r e^{-x} dx = \int_0^\infty \frac{e^{-x} dx }{1-qx}
 \label{factorialelimination}
\ee
Now, it is clear that the r.h.s. depends on the integration contour,
and there is an ambiguity
\be
\oint_{x=q^{-1}} \frac{e^{-x} dx }{1-qx} = \frac{2\pi i}{q} e^{-{1/q}}
\ee
so that the sum of divergent series is defined modulo this ``instantonic" term.

This standard argument, lying in the base of entire "resurgence theory" \cite{resurgence}
is, however, technically based on exponential functions, and their Gamma and,
more general, hypergeometric generalizations.
They have, of course, a straightforward $q$ (``quantum") and $t$ deformations \cite{GR,Mac1,Koor},
but this is not the only important direction to look at.
In modern integrability theory \cite{UFN3}, which is a crucially important part of
non-perturbative physics, an
important role is played by {\it combinatorial} generalizations of the factorial,
like the quantity $d_R$, which extends the factorial from integers to partitions (Young diagrams).
It seems important to look at the versions of Borel transform that deal with
such additional combinatorial structure, hence, the name combinatorial Borel transform.
Surprisingly or not, exponentials still show up in the combinatorial Borel transforms, through Cauchy  identities
\cite{Mac}, thus most probably this is not the last step.
However, it can prove sufficient for the needs of integrability theory, at least.

As to possible techniques which could provide a route to the combinatorial Borel transforms,
in  \cite{2212.04859}  it was suggested to consider an interpolation between
two different kinds of sums over all Young diagrams
\be
\sum_R q^{|R|} = \prod_{k=1}\frac{1}{1-q^k}
\label{gendia}
\ee
and
\be
\sum_R q^{|R|} S_R\{p\}S_R\{\bar p\} = \exp\left(\sum_k \frac{q^kp_k\bar p_k}{k}\right)
\label{Cauchy}
\ee
by taking a specially adjusted average of the Schur functions over the power sums $p_k$.
The emphasis, however, can be put not so much on averaging, but on coexistence
of generating functions differing by the non-trivial combinatorial weights $d_R$.
It is a common place to consider such families differing by the number of factorials
in the weights: these are just related by the standard exponential Borel transform.
It is an interesting direction to study a more tricky combinatorial difference.
The goal of the present paper is to
exploit this idea and extend it further in various directions.

\begin{figure}
\begin{picture}(300,100)(-50,-70)
\put(0,0){\mbox{$\sum_r a_r q^r$}}
\put(14,-10){\vector(0,-1){25}}   \put(-25,-25){\mbox{\text{  Borel  }}}
\put(18,-35){\vector(0,1){25}}   \put(24,-21){\mbox{\footnotesize\text{ $\Gamma$-average  }}}
\put(22,-31){\mbox{\footnotesize $\left<q^r\right>=q^r \frac{a_r}{b_r}$ }}
\put(0,-50){\mbox{$\sum_r b_r q^r$}}
\put(60,0){\mbox{$=$}}
\put(60,-50){\mbox{$=$}}
\put(45,-60){\mbox{\footnotesize $S_r\{0\}=\frac{b_r}{c_r}$}}
\put(105,0){\mbox{$\sum_r a_r q^r $}}   
\put(100,-50){\mbox{$\underline{\sum_r c_r q^r S_r\{\bar p\}}$}}
\put(120,-37){\vector(0,1){26}}
\put(125,-25){\mbox{\footnotesize $\left<S_r\{\bar p\}\right>_{\bar p} =\frac{a_r}{c_r}$}}
\put(185,5){\vector(1,0){40}}
\put(185,8){\vector(1,0){40}}
\put(185,-45){\vector(1,0){40}}
\put(185,-48){\vector(1,0){40}}
\put(295,-25){\vector(0,1){15}} \put(299,-10){\vector(0,-1){17}}
\put(260,-22){\mbox{\text{  CBT  }}}
\put(307,-22){\mbox{$\left<S_R\{\bar p\}\right>_{\bar p} = \frac{a_R}{c_R}$}}
\put(250, 0){\mbox{$\sum_R a_R q^{|R|} S_R\{p\} =F\{p\}$}}
\put(250,-50){\mbox{$\boxed{\sum_R c_R q^{|R|} S_R\{p\} S_R\{\bar p\}}$}}
\put(175,20){\mbox{\text{generalization}}}
\end{picture}
\caption{\footnotesize  A picture, illustrating the concept of the {\it combinatorial Borel transform}
(CBT) of the series.
First, we introduce into the original series an auxiliary function $S_r\{\bar p\}$
such that its average over $\bar p$ with an appropriate measure can change the coefficients
in a desired way: say, $a_r=r!\longrightarrow b_r=1$ in (\ref{factorialelimination}) or,
more general, $a_r=d_r^m = (r!)^{-m} \longrightarrow b_r =a_rd_r = d_r^{m+1}=(r!)^{-m-1}$.
For traditional applications, the main point is that the underlined sum can converge better
than the original one in the left upper corner due to
$S_r\{0\} = \frac{b_r}{c_r} \sim \frac{1}{r!}$,
but for us this is just one application of many.
Then, we extend the summation domain $\{r\}\longrightarrow \{R\}$
and introduce an additional auxiliary function $S_R\{p\}$
so that the sum in the box is easily calculated, for instance, with the help of the Cauchy identity.
As an additional bonus, this converts the Cauchy relations into ineteresting formulas for the series
$F\{p\}$ at the right upper corner, which can  {\it a priori} look non-trivial,
but are actually the $\bar p$ averages of simple identities.
Technically, what we do is a substitution of the Gamma-function average
$\left<q^r\right> \longrightarrow q^r\Gamma(r+1) = r!\cdot q^r$
which inverts the left vertical arrow in the usual Borel transform, by a smoother and richer
Gaussian average, at the price of extending the sum over integers $r$ to that over all integer
partitions (Young diagrams) $R$.
Such generalization is not unique, which opens a way to obtain a whole class of
formulas by slight changes of the Gaussian measure.
}
\end{figure}

The paper is organized as follows. In section 2, we explain the basic of our approach, and demonstrate how the averaging method works for evaluating sums. In section 3, we extend the method to reproduce multiple combinatorial sums. In section 4, we discuss evaluating bilinear averages in parallel with the property of strong superintegrability in matrix models. Section 5 contains short notes on relations of the sums considered in the paper with the Ramanujan sums. At last, section 6 discusses the cases when the sums are divergent, while section 7 contains some concluding remarks.

\paragraph{Notation.}
We use the notation $S_R\{p_k\}$ for the Schur functions, which are symmetric functions of $x_i$, and are graded polynomials of power sums $p_k:=\sum_ix_i^k$. They are labeled by partitions, or, equivalently, by Young diagrams $R$ with $l_R$ parts: $R_1\ge R_2\ge\ldots\ge R_{l_R}$.
The corresponding skew Schur functions labeled by two partitions $R$ and $Q\in R$ are denoted as $S_{R/Q}\{p_k\}$, while the Macdonald polynomials are denoted through $M_R\{p_k\}$. We also denote $R^\vee$ the conjugate partition (Young diagram).

Our main quantity in this paper is $d_R$,
\be
d_R = \frac{\prod_{i<j}^{l_R} \left(R_i -i - R_j+j\right)}
{\prod_{i=1}^{l_R} \left(l_R +R_i-i   \right)!}
\ee
which is equal to $S_R\{\delta_{k,1}\}$ in terms of Schur functions, and is equal to ${\rm dim}_R/|R|!$, where ${\rm dim}_R$ is the dimension of representation $R$ of the symmetric group ${\cal S}_{|R|}$. $|R|=\sum_i^{l_R}R_i$ is here the size of partition $R$ with $l_R$ parts.

The theory of symmetric functions can be found in \cite{Mac}, and the theory of symmetric groups, in \cite{Fulton}.

We also use notation $\sum_R$ for the sum over all partitions $R$ including the empty set, while $\sum_{R\vdash m}$ means summing over all partitions of integer $m$.

\bigskip

\section{Basic example of averaging method}

Eq.(\ref{gendia}) counts Young diagrams, or the ordered integer partitions,
while (\ref{Cauchy}) is the Cauchy identity for the Schur functions, i.e. for the characters of
linear groups $GL_N$.
Technically, upon choosing $p_k=\bar p_k=\delta_{k,1}$ in (\ref{Cauchy}), the difference between the two formulas is in the power of
$d_R:=S_R\{\delta_{k,1}\}$: they are the sums of $d_R^0$ and $d_R^2$, accordingly.
Generally, $|R|!\cdot d_R={\rm dim}_R$ are {\it integer}-valued
dimensions of representations of the symmetric group
${\cal S}_{|R|}$ \cite{Fulton},  and the sum
\be
\xi_m:=\sum_R q^{|R|} d_R^m = \sum_R q^{|R|}\frac{{\rm dim}_R^m}{(|R|!)^m}
\label{dimrep}
\ee
looks somewhat terrible. The first terms in (\ref{dimrep}) are
\be
\xi_m = 1 + q + \frac{2}{(2!)^m} q^2
+ \frac{2^m+2}{(3!)^m} q^3
+ \frac{2\cdot 3^m+2^m+2}{(4!)^m} q^4
+ \frac{6^m+2\cdot 5^m +2\cdot 4^m+2}{(5!)^m} q^5 +
\nn \\
+ \frac{16^m+2\cdot 10^m+2\cdot 9^m+4\cdot 5^m+2}{(6!)^m} q^6
+ \frac{2\cdot 35^m+2\cdot 21^m + 20^m +2\cdot 15^m +4\cdot 14^m+2\cdot 6^m+2}{(7!)^m} q^7 +
\nn \\
+ \frac{90^m + 2\cdot 70^m+2\cdot 64^m+2\cdot 56^m+42^m + 2\cdot 35^m +2\cdot 28^m +2\cdot 21^m + 2\cdot 20^m
+ 2\cdot 14^m+2\cdot 7^m+2}{(8!)^m} q^8 +
\nn
\ee
\be
+ \ldots
\label{xim}
\ee
The more surprising is that, at some particular values of $m$, one gets sensible formulas.

The best known is the quadratic case, $m=2$, where the generic theory of finite groups
implies that the sum of the squared dimensions is always a dimension of the group.
In the case of ${\cal S}_r$, this gives
\be
\sum_{R\vdash r}  {\rm dim}_R^2 = r!
\ee
and leads to the particular case of (\ref{Cauchy})
\be
\xi_2=
\sum_R q^{|R|} S_R\{\delta_{k,1}\}^2 = \sum_R q^{|R|} d_R^2 = e^{q}
\label{0powerdR}
\ee

For $m=0$, one gets a far more transcendental, still comprehensible formula (\ref{gendia}),
which just counts the numbers of representations of ${\cal S}_{|R|}$.

Amusingly, the case of $m=1$ is also simple:
\be
\Sigma_n=\sum_{R\vdash n}q^n {\rm dim}_R=n!\sum_{R\vdash n}q^n d_R
\ee
gives rise to sequence A000085 from \cite{seqs} which counts the numbers of Young {\it tableaux}
of a given size $r$, and this is in perfect accordance with the fact that the dimension
${\rm dim}_R$ is equal to the number of Young {\it tableaux} associated with the diagram $R$.  For the exponential generating function of this sequence, see \cite[sequence A000085 and references therein]{seqs},
\be
\xi_1 =
\sum_R q^{|R|}d_R = \sum_{n=0}^\infty{\Sigma_n\over n!} = e^{q+\frac{q^2}{2}}
\label{1powerdR}
\ee
Moreover, this sum can be lifted to a sum of single(!) Schur functions
\be
\boxed{
\sum_R q^{|R|}  S_R\{p\} =
\sum_R q^{|R|}  S_R\{p\} \underbrace{\left<S_R\{\bar p\}\right>^{\bf\!A}_{\!\bar p}}_{1}
\ \stackrel{(\ref{Cauchy})}{=}\
\left<\exp\left(\sum_{k=1}^\infty\frac{q^kp_k\bar p_k}{k}\right)\right>^{\bf\!A}_{\bar p}=
\exp\left\{\sum_{k=1}^\infty \left(q^{2k}\frac{p_k^2}{2k}
+ q^{2k-1}\frac{p_{2k-1}}{2k-1}\right)\right\}
}
\label{idlinSchur}
\ee
In the language of \cite{2212.04859}, this follows from taking average of the Cauchy identity
(\ref{Cauchy}) over the variables $\bar p_k$ with the measure
\be\label{Gm}
\text{\bf Measure A:} \ \ \ \ \ \ \ \ \
\Big<S_R\Big>^{\bf\!A}=1: && \ \ \ \ \ \ \
d\mu^{\bf A} \sim \prod_{k=1}^\infty dp_k\exp\left(-\frac{(p_{2k-1}-1)^2}{2(2k-1)} - \frac{p_{2k}^2}{4k}\right)
\ee
Hereafter, the average with a given measure $d\mu$ over variables $p_k$, $k=1,2,\ldots$ is defined as
\be\label{average}
\Big< f(\{p_k\})\Big>:=\int_{-\infty}^{\infty} d\mu f(\{p_k\})
\ee
and normalized in such a way that $\Big<1\Big>=1$.

Average (\ref{Gm}) converts the Schur polynomials into 1, $\left<\bar S_R\{\bar p\}\right>^{\bf\!A}=1$,
and is Gaussian, so it is trivial to take averages of the r.h.s. of (\ref{Cauchy}),
\be
\left<\exp\left(\sum_{k=1}^\infty\frac{q^k p_k\bar p_k}{k}\right)\right>^{\bf\!A}_{\!\bar p}
\sim \prod_{k=1}^\infty \left(\int e^{\frac{q^{2k-1} p_{2k-1}\bar p_{2k-1}}{2k-1}}
e^{-\frac{(\bar p_{2k-1}-1)^2}{2(2k-1)}} d\bar p_{2k-1}
\int e^{\frac{q^{2k} p_{2k}\bar p_{2k}}{2k}}
e^{-\frac{\bar p_{2k}^2}{4k}} d\bar p_{2k}
\right)
\ee
is just the r.h.s. of   (\ref{idlinSchur}).

The sum (\ref{1powerdR}) can be also represented as
\be
\sum_R q^{|R|}d_R = \sum_R q^{R} S_R\{\delta_{k,1}\}\underbrace{\left<S_R\{\bar p_k\}\right>^{\bf\!A}_{\bar p}}_{1}
\ \stackrel{(\ref{Cauchy})}{=} \
\left<e^{ q\bar p_1}\right>^{\bf\!A} = \sum_{r=0}^\infty \frac{q^r \!\left<\bar p_1^r\right>^{\bf\!A}}{r!}
\ee
i.e.
\be
r!\cdot \sum_{R\vdash r}  d_R = \left< p_1^r\right>^{\bf\!A}
\ee

Note that while (\ref{gendia}) is a generating function for numbers of Young diagrams,
(\ref{1powerdR}) is an {\it exponential} (i.e. weighted with additional factors $\frac{1}{r!}$) generating function for numbers of Young {\it tableaux}.

Another important difference between counting diagrams and tables is that the latter,
being related to dimensions, admit a quantum deformation, see sec.\ref{Mac} below.

According to \cite{seqs}, no interpretations for $\xi_m$ are yet known for $m\geq 3$.

\section{Simple generalizations}

\subsection{Derivation of various sum formulas by the averaging method}

The averaging method can be easily modified to handle more examples.
Say, we can ask the average $\left<S_R\right>=1$ only for {\it even} diagrams $R$,
i.e. with all rows of even lengths, and vanishes otherwise.
The corresponding measure is still Gaussian,
\be
\text{\bf Measure B}: \ \ \ \ \ \ \ \ \ \ \
\left\{\begin{array}{ccc}\left<S_R\right>^{\bf\!B}=1 && {\rm for\ even}\ R  \\
\left<S_R\right>^{\bf\!B}=0 && {\rm otherwise} \end{array} \right. :  \ \ \ \ \ \ \ \
d\mu^{\bf\!B} \sim \prod_{k=1}^\infty dp_k \exp\left(-\frac{p_{2k-1}^2}{2(2k-1)} - \frac{(p_{2k}-1)^2}{4k}\right)
\ee
and the counterpart of (\ref{idlinSchur}) is
\be
\sum_{R\ even}S_R
= \sum_R S_R\underbrace{\left<S_R\right>^{\bf\!B}}_{\delta_{R,{\rm even}}}
= \left<e^{\sum_{k=1}^\infty \frac{p_k\bar p_k}{k}}\right>^{\bf\!B}_{\bar p}
= \prod_{k=1}^\infty \exp\left({p_k^2\over 2k}+{p_{2k}\over 2k}\right)
\ee
Let us demonstrate how to prove that $\left<S_R\right>^{\bf\!B}$ is non-zero and equal to 1 only for even partitions. To this end, we note that the measure ${\bf\!B}$ is obtained from ${\bf\!A}$ by the shift of times $p_k\to p_k-(-1)^k$. In other words, we have
\be
\left<S_R\{p_k\}\right>^{\bf\!B}=\left<S_R\{p_k+(-1)^k\}\right>^{\bf\!A}=\sum_P
S_{R/P}\{(-1)^k\}\left<S_R\{p_k\}\right>^{\bf\!A}\stackrel{[8]}{=}\sum_P S_{R/P}\{(-1)^k\}=\sum_P
(-1)^{|R|+|P|}S_{R/P}\{1\}
\ee
The choice of $p_k=1$ is equivalent to choosing only one non-zero symmetric variable $x_1=1$, and the sum $\sum_P(-1)^{|P|}S_{R/P}\{1\}$ is not zero only when $R$ is an even partition, which follows from the representation of the skew Schur function $S_{R/P}=\sum_Tx^T$ with the sum running over all tableaux $T$ of shape $R-P$, see details in \cite[Eq.(5.12)]{Mac}.

All further examples in this section are proved analogously. Note that {\bf checking} this kind of claims is much easier than proving them: one can just calculate simple Gaussian averages of the Schur functions at any concrete example.

Now, in order to pick up only even {\it columns}, one need to make transposition,
which is equivalent to changing signs of all the time variables:
the relevant measure, which gives $\left<S_R\right>=1$ for all even-column diagrams and zero otherwise is
\be
\text{\bf Measure C:} \ \ \ \ \ \ \ \
\left\{\begin{array}{ccc}\left<S_R\right>^{\!C}=1 && {\rm for\ even} \ R^\vee  \\
\left<S_R\right>^{\!C}=0 && {\rm otherwise} \end{array} \right. :  \ \ \ \ \ \ \ \ \ \ \
d\mu^{\bf\!C} \sim \prod_{k=1}^\infty dp_k\exp\left(-\frac{p_{2k-1}^2}{2(2k-1)} - \frac{(p_{2k}+1)^2}{4k}\right)
\ee
and
\be
\sum_{R^\vee\ even}S_R =  \sum_R S_R\underbrace{\left<S_R\right>^{\bf\!C}}_{\delta_{R^\vee,{\rm even}}}
= \left<e^{\sum_{k=1}^\infty \frac{p_k\bar p_k}{k}}\right>^{\bf\!C}_{\bar p}
= \prod_{k=1}^\infty  \exp\left({p_k^2\over 2k}-{p_{2k}\over 2k}\right)
\ee
Likewise the average with $\left<S_R\right>^{\bf \!D} = (-1)^{\nu_R}(-1)^{3|R|/2}$
with $\nu_R = \sum_{i=1}^{l_R} (i-1)R_i$
is provided by the Gaussian measure
\be
\text{\bf Measure D:} \ \ \
\left<S_R\right>^{\bf \!D} = (-1)^{\nu_R}(-1)^{3|R|/2}: \ \ \ \
d\mu^{\bf\!D} \sim \prod_{k=1}^\infty \exp\left(-\frac{p_{k}^2}{2k} + (-1)^k\sqrt{-1}\frac{p_{2k-1}}{2k-1}
- \frac{p_{4k-2}}{2k-1}\right) dp_k
\label{muexpcaj}
\ee
Then
\be
\sum_R (-1)^{\nu_R} S_R\{p\} = \sum_R (-1)^{|R|/2} S_R\{p\} \left<S_R\right>^{\bf\!D}
&=& \left<\exp\left(\sum_{k=1}^\infty(-1)^{k/2}\frac{ p_k\bar p_k}{k}\right)\right>^{\bf \!D}_{\!\!\bar p} =\nn\\
&=&\prod_{k=1}^\infty \exp\left(-\frac{p_{2k-1}^2}{2(2k-1)} + \frac{p_{2k}^2}{4k} + \frac{p_{2k-1}}{2k-1}
 + \frac{p_{4k-2}}{2k-1}\right)
\ee
If $c_R$ counts the number of rows of odd length in $R$,
then, introducing a formal parameter $t$,
\be
\text{\bf Measure E:} \ \ \ \ \
\left<S_R\right>^{\bf \!E} = t^{c_R}: \ \ \ \ \ \ \ \ \ \ \
d\mu^{\bf \!E} \sim \prod_{k=1}^\infty \exp\left(-\frac{(p_{2k-1}-t^{2k-1})^2}{2(2k-1)} -
\frac{(p_{2k}-1+t^{2k})^2}{4k}\right) dp_k
\ee
and
\be
\sum_Rt^{c_R}S_R=\prod_i{1\over 1-tx_i}\prod_{i<j}{1\over 1-x_ix_j}
=\exp\left\{\sum_{k=1}^\infty\left({p_k^2\over 2k}+t^{2k-1}{p_{2k-1}\over 2k-1}+(t^{2k}-1){p_{2k}\over 2k}\right)\right\}
\ee
Counting the number of columns of odd length is provided by the same formula
with all the signs of $p_k$ inverted, see (\ref{1}) below.

\subsection{Extension to skew functions}

Now, one can easily obtain similar formulas for sums of the skew Schur functions $S_{R/P}$.
To this end, one needs a slight generalization of the Cauchy identity (\ref{Cauchy}):
\be\label{C2}
\sum_R S_{R/P}\{p\}S_{R}\{\bar p\}=\exp\left(\sum_{k=1}^\infty{p_k\bar p_k\over k}\right)S_{P}\{\bar p\}
\ee
Taking the average of this identity over $\bar p_k$ with measure (\ref{Gm}), one obtains
\be\label{av}
\sum_R S_{R/P}\{p\}\Big<S_{R}\{\bar p\}\Big>^{\bf\!A}_{\bar p}=\sum_R S_{R/P}\{p\}=
\left<\exp\left(\sum_{k=1}^\infty{p_k\bar p_k\over k}\right)S_{P}\{\bar p\}\right>^{\bf\!A}_{\bar p}
\ee
Let us prove that the r.h.s. of this formula is given by
\be\label{pr}
\left<\exp\left(\sum_{k=1}^\infty{p_k\bar p_k\over k}\right)S_{P}\{\bar p\}\right>^{\bf \!A}_{\!\!\bar p}=
\exp\left\{\sum_{k=1}^\infty \left(\frac{p_k^2}{2k}+ \frac{p_{2k-1}}{2k-1}\right)\right\}\sum_QS_{P/Q}\{p\}
\ee
with literally the same $p$-dependent factor as in (\ref{idlinSchur}).
To this end, we convert this formula with the Schur functions $S_P\{p'\}$.
This gives
\be
\left<\exp\left(\sum_{k=1}^\infty{(p_k+p'_k)\bar p_k\over k}\right)\right>^{\bf\!A}_{\!\!\bar p}=
\exp\left\{\sum_{k=1}^\infty \left(\frac{p_k^2}{2k}+ \frac{p_{2k-1}}{2k-1}+{p_kp'_k\over k}\right)\right\}\sum_QS_Q\{p'\}
\ee
where we once again used the Cauchy identities (\ref{Cauchy}) and (\ref{C2}) at the l.h.s. and at the r.h.s.
Now it remains to calculate the Gaussian integral at the l.h.s. as it was done in (\ref{idlinSchur}),
and also to use (\ref{idlinSchur}) to evaluate the sum at the r.h.s.
Hence, formula (\ref{pr}) is correct, and we finally obtain from (\ref{av})
\be\label{sS}
\sum_R S_{R/P}\{p\}=\exp\left\{\sum_{k=1}^\infty \left(\frac{p_k^2}{2k}+ \frac{p_{2k-1}}{2k-1}\right)\right\}\sum_QS_{P/Q}\{p\}
\ee
We can now make one more check that (\ref{sS}) is correct.
Convert this formula with $S_P\{p'\}$ and use the Cauchy identity (\ref{C2}) and the definition of the skew Schur polynomial
\be
S_R\{p+p'\}=\sum_PS_{R/P}\{p\}S_P\{p'\}
\ee
Then
\be
\sum_RS_R\{p+p'\}=\exp\left\{\sum_{k=1}^\infty \left(\frac{p_k^2}{2k}+ \frac{p_{2k-1}}{2k-1}+{p_kp'_k\over k}\right)\right\}
\sum_QS_Q\{p\}
\ee
Both sums can be calculated using (\ref{idlinSchur}) in order to validate this formula.
We emphasize once again that, according to this argument, the $p$-dependent factor at the r.h.s.
is literally the same as in (\ref{idlinSchur}),
and the same will be true for the skew Schur counterparts of all other identities in this section.

\subsection{A collection of sum formulas\label{Mac}}

To summarize, we can reproduce the whole collection of the single-Schur sums from \cite[secs.1.5,3.4,3.5]{Mac}.

\bigskip

They are naturally divided in three blocks, and the formulas are given also in terms of symmetric variables $x_i$ such that $p_k=\sum_ix_i^k$.

\bigskip

{\bf 1.} The first one is already familiar
\be\label{1}
\sum_RS_R=\prod_i{1\over 1-x_i}\prod_{i<j}{1\over 1-x_ix_j}
= \exp\left\{\sum_{k=1}^\infty \left({p_k^2\over 2k}+{p_{2k-1}\over 2k-1}\right)\right\}\\
\sum_{R\ even}S_R=\prod_i{1\over 1-x_i^2}\prod_{i<j}{1\over 1-x_ix_j}
=\exp\left\{\sum_{k=1}^\infty\left({p_k^2\over 2k}+{p_{2k}\over 2k}\right)\right\}\\
\sum_{R^\vee\ even}S_R=\prod_{i<j}{1\over 1-x_ix_j}=
\exp\left\{\sum_{k=1}^\infty\left({p_k^2\over 2k}-{p_{2k}\over 2k}\right)\right\}\\
\sum_R(-1)^{\nu_R}S_R=\prod_i{1\over 1-x_i}\prod_{i<j}{1\over 1+x_ix_j}=
\exp\left\{\sum_{k=1}^\infty\left((-1)^k{p_k^2\over 2k}+{p_{2k-1}\over 2k-1}+{p_{4k-2}\over 2k-1}\right)\right\}\\
\sum_Rt^{c_R}S_R=\prod_i{1\over 1-tx_i}\prod_{i<j}{1\over 1-x_ix_j}
=\exp\left\{\sum_{k=1}^\infty\left({p_k^2\over 2k}+t^{2k-1}{p_{2k-1}\over 2k-1}+(t^{2k}-1){p_{2k}\over 2k}\right)\right\}\\
\sum_Rt^{r_R}S_R=\prod_i{1+tx_i\over 1-x_i^2}\prod_{i<j}{1\over 1-x_ix_j}
=\exp\left\{\sum_{k=1}^\infty\left({p_k^2\over 2k}+t^{2k-1}{p_{2k-1}\over 2k-1}+(1-t^{2k}){p_{2k}\over 2k}\right)\right\}
\ee
where $c_R$ ($r_R$) is the number of rows (lines) of odd length, and $\nu_R=\sum_i(i-1)R_i$. The first three lines follow from the last two lines upon putting $t=0,1$.

\bigskip

{\bf 2.} The second block involves the skew Schur functions:
\be
\sum_RS_{R/Q}=\prod_i{1\over 1-x_i}\prod_{i<j}{1\over 1-x_ix_j}\sum_PS_{Q/P}
=\exp\left\{\sum_{k=1}^\infty\left({p_k^2\over 2k}+{p_{2k-1}\over 2k-1}\right)\right\}\sum_PS_{Q/P}\\
\sum_{R\ even}S_{R/Q}=\prod_i{1\over 1-x_i^2}\prod_{i<j}{1\over 1-x_ix_j}\sum_{P\ even}S_{Q/P}
=\exp\left\{\sum_{k=1}^\infty\left({p_k^2\over 2k}+{p_{2k}\over 2k}\right)\right\}\sum_{P\ even}S_{Q/P}\\
\sum_{R^\vee\ even}S_{R/Q}=\prod_{i<j}{1\over 1-x_ix_j}\sum_{P^\vee\ even}S_{Q/P}
=\exp\left\{\sum_{k=1}^\infty\left({p_k^2\over 2k}-{p_{2k}\over 2k}\right)\right\}\sum_{P^\vee\ even}S_{Q/P}
\ee
for arbitrary $Q$.

\bigskip

{\bf 3.} The third direction is generalizations of these formulas to the Macdonald polynomials.
We do not provide relevant averages  for this case in order to avoid overloading the text
with unnecessary complicated formulas.
Still, we list the results.
Denote
\be
H_R^{el}:=\prod_{{(i,j)\in R}\atop{R_j^\vee-i\ even}}{1-q^{R_i-j}t^{R_j^\vee-i+1}\over 1-q^{R_i-j+1}t^{R_j^\vee-i}},
\ \ \ \ \ H_R^{oa}:=\prod_{{(i,j)\in R}\atop{R_i-j\ odd}}{1-q^{R_i-j}t^{R_j^\vee-i+1}\over 1-q^{R_i-j+1}t^{R_j^\vee-i}}
\ee
The full product $H_R$ is just the coefficient that stands in the sum of the Cauchy identity:
\be
\sum_RH_RM_R\{p_k\}M_R\{\bar p_k\}=
\prod_{i,j}{(tx_i\bar x_j;q)_\infty\over (x_i\bar x_j;q)_\infty}=\exp\left\{\sum_{k=1}^\infty\left({1-t^k\over 1-q^k}{p_k\bar p_k\over k}\right)
\right\}
\ee
with $p_k=\sum_ix_i^k$, $\bar p_k=\sum_i\bar x_i^k$.
Then,
\be
\sum_RH_R^{el}M_R=\prod_i{(tx_i;q)_\infty\over (x_i;q)_\infty}\prod_{i<j}{(tx_ix_j;q)_\infty\over (x_ix_j;q)_\infty}=
\exp\left\{\sum_{k=1}^\infty {1-t^k\over 1-q^k}\left({p_k^2\over 2k}+{p_{2k-1}\over 2k-1}\right)\right\}\\
\sum_RH_R^{oa}M_R=\prod_i{(qtx_i^2;q^2)_\infty\over (1-x_i)(q^2x_i^2;q^2)_\infty}
\prod_{i<j}{(tx_ix_j;q)_\infty\over (x_ix_j;q)_\infty}=
\exp\left\{\sum_{k=1}^\infty\left({1-t^k\over 1-q^k}{p_k^2\over 2k}+{t^k-q^k\over 1+q^k}{p_{2k}\over 2k}+{p_{2k-1}\over 2k-1}\right)
\right\}\\
\sum_{R^\vee\ even}H_R^{el}M_R=\prod_{i<j}{(tx_ix_j;q)_\infty\over (x_ix_j;q)_\infty}=
\exp\left\{\sum_{k=1}^\infty {1-t^k\over 1-q^k}\left({p_k^2\over 2k}-{p_{2k}\over 2k}\right)\right\}\\
\sum_{R\ even}H_R^{oa}M_R=\prod_i{(qtx_i^2;q^2)_\infty\over (x_i^2;q^2)_\infty}
\prod_{i<j}{(tx_ix_j;q)_\infty\over (x_ix_j;q)_\infty}=\prod_{i<j}{(tx_ix_j;q)_\infty\over (x_ix_j;q)_\infty}=
\exp\left\{\sum_{k=1}^\infty\left({1-t^k\over 1-q^k}{p_k^2\over 2k}+{1+t^k\over 1+q^k}{p_{2k}\over 2k}\right)\right\}
\ee

\subsection{The Gaussian measure}

The main Gaussian measure (\ref{Gm})
itself enjoys a curious property: it can be expanded into a simple sum
over the Schur functions containing only self-conjugated Young diagrams:
\be\label{Gmex}
d\mu \sim
\exp\left\{-\sum_{k=1}^\infty \left( \frac{p_{2k-1}^2-2p_{2k-1}}{2(2k-1)} + \frac{p_{2k}^2}{4k}\right)\right\}
=\sum_{R=R^\vee}(-1)^{|R|(|R|+1)+h_R(h_R+1)\over 2}S_R
= \nn \\
= \sum_{h_R=1}^\infty \sum_{k_1\ge k_2 \ge k_{h_R}\geq 0}^\infty  (-1)^{\sum_{i=1}^{h_R} k_i}\cdot
S_{(k_1,\ldots,k_{h_R}|k_1,\ldots,k_{h_R})}
\ee
where $h_R$ is the number of hooks that form the Young diagram $R$, and $(\vec \alpha|\vec\beta)$ is the Fr\"obenius notation of the Young diagram \cite{Mac}.

\subsection{Calculating various sums}

Note that the averaging procedure allows one to evaluate various sums of combinatorics quantities by evaluating the Gaussian moment. For instance (using the Fr\"obenius formula \cite{Fulton}),
\be
\left<p_\Delta\right>^{\bf A}=\sum_R\psi_R(\Delta)\left<S_R\right>^{\bf A}=\sum_R\psi_R(\Delta)
\ee
i.e. (see also \cite[Example 11, sec.I.7]{Mac})
\be
\sum_R\psi_R(\Delta)=\prod_{i=1}^\infty\left<p_i^{m_i}\right>^{\bf A}
\ee
are products of the Gaussian moments. Here $\Delta$ is a partition with parts $\delta_i$, $p_\Delta:=\prod_{i=1}^{l_\Delta}p_{\delta_i}$, and $\psi_R(\Delta)$ is the character of the symmetric group ${\cal S}_{|R|}$ in the representation $R$.
Also
\be
\sum_PN^R_{PQ}=\left<S_{R/Q}\right>^{\bf A}\\
\sum_RN^R_{PQ}=\left<S_PS_Q\right>^{\bf A}\\
\hbox{etc.}
\ee
where $N^R_{PQ}$ are the Littlewood-Richardson coefficients,
\be
S_PS_Q=\sum_RN^R_{PQ}S_R
\ee
and we used that \cite{Mac}
\be
S_{R/Q}=\sum_PN^R_{PQ}S_P
\ee

\section{Strong superintegrability and $W$-operators}

As we saw in the previous sections, the Schur functions form a full set of polynomials that have simple averages, this phenomenon is called superintegrability \cite{MMsi}. Similarly, one can construct a full set of {\it bilinear} combinations with simple averages too.
This phenomenon is called strong superintegrability, and first we remind how this phenomenon looks like in the case of the Gaussian
Hermitian matrix model.

\subsection{Strong integrability in Gaussian Hermitian matrix model}

The bilinear correlators in the Gaussian Hermitian matrix model are generated by the action of the $W$-operators $\hat W^{(-)}$ on the Schur function $S_R$ as functions of $P_k:=\Tr H^k$ \cite{MMNek},
\be\label{da}
\Big<S_Q\{\hat W^{(-)}_k\}\cdot S_R\{P_k\}\Big>=
{\displaystyle{S_{R/Q}\{\delta_{k,2}\}S_R\{N\}}\over \displaystyle{S_R\{\delta_{k,1}\}}}\nn\\
\hat W^{(-)}_k:=\Tr \left({\p\over\p H}\right)^k
\ee
where $H$ is the matrix that is integrated over in the matrix model, and, by the matrix derivative, we imply the derivative w.r.t. matrix elements of the transposed matrix: $\left(\frac{\partial}{\partial H}\right)_{ij}=\frac{\partial}{\partial H_{ji}}$.

As we demonstrated in \cite{MMd}, the averages (\ref{da}) can be reduced to a correlator of the form
\be
\Big<S_Q\{\hat W^{(-)}_k\}\cdot S_R\{P_k\}\Big>=\Big<K_Q\{P_k\}\cdot S_R\{P_k\}\Big>
\ee
where the polynomials $K_R$ form a complete basis, and celebrate the property
\be
\Big<K_R\cdot K_Q\Big>=\displaystyle{S_{R}\{N\}\over \displaystyle{S_R\{\delta_{k,1}\}}}\delta_{RQ}
\ee
Examples of these polynomials can be found in \cite[Appendix]{MMd}. They are obtained by using integration by parts and the action of the $S_Q\{\hat W^{(-)}_k\}$-operators on the Gaussian measure.

Note that, throughout the paper \cite{MMd}, we discussed another basis of polynomials, $K_\Delta$, the two being related by the Fr\"obenius formula
\be
K_R=\sum_\Delta{\psi_R(\Delta)\over z_\Delta}K_\Delta\nn
\ee
where $z_\Delta$ is the standard symmetric factor of the Young diagram (order of the automorphism) \cite{Fulton}.

\subsection{Strong superintegrability in the $p$-Gaussian model}

In the $p$-Gaussian model with the measure (\ref{Gm}), which is under our consideration in this paper, there is a formula for the bilinear correlator similar to (\ref{da}), however, in this case, one has to deal with the operators\footnote{The superscript ``0" here refers to a commutative part of the $w_{\infty}$ algebra, while the superscript ``-" in the previous subsection to its Borel part.} $\hat W_\Delta^{(0)}$. They are defined as
\be
\hat W_\Delta^{(0)}={1\over z_\Delta}\ :\ \prod_{i=1}^{l_\Delta}\Tr \left(\Lambda{\p\over\p\Lambda}\right)^{\delta_i}\ :
\ee
where $\Lambda$ is a matrix such that the $p_k$-variables are parameterized as $p_k=\Tr \Lambda^k$, and the invariant operators $\hat W_\Delta^{(0)}$ can be rewritten as operators in $p_k$-variables. The normal ordering here, $:\ldots:$ means that all the derivatives are placed to the rightmost positions.

Note that these operators $\hat W_\Delta^{(0)}$ require the normal ordering, otherwise they would depend on the size of the $\Lambda$ matrix \cite{MMN}. In the case of operators $\hat W_\Delta^{(-)}$ in (\ref{da}), normal ordering did not affect the answer. Note that the Schur functions form a set of eigenfunctions of $\hat W_\Delta^{(0)}$:
\be
\hat W_\Delta^{(0)}S_R=\phi_R(\Delta)S_R
\ee
where the eigenvalue $\phi_R(\Delta)$ is manifestly expressed through the symmetric group characters $\psi_\mu(\Delta)$ by \cite{IK,Ok,MMNspin}
\be
\phi_R(\Delta)=\sum_{\mu\vdash\Delta} {S_{R/\mu}\{\delta_{k,1}\}\over S_R\{\delta_{k,1}\}}{\psi_\mu(\Delta)\over z_\Delta}
\ee
Now using the identity
\be
\sum_{\Delta\vdash P}\psi_P(\Delta)\phi_R(\Delta)={S_{R/P}\{\delta_{k,1}\}\over S_{R}\{\delta_{k,1}\}}
\ee
which follows from the orthogonality relation
\be
\sum_\Delta{\psi_P(\Delta)\psi_R(\Delta)\over z_\Delta}=\delta_{PR}
\ee
one finally comes to the formula
\be\boxed{
\Big<S_P\{\hat W^{(-)}_k\}\cdot S_R\{p_k\}\Big>^{\bf A}={S_{R/P}\{\delta_{k,1}\}\over S_{R}\{\delta_{k,1}\}}
}
\ee

Now, using integration by parts, one could  again recast the action of the $S_P(\hat W^{(0)}_k)$-operators into the action to the measure (\ref{Gm}) in order to produce a complete set of polynomials $K_P\{p_k\}$ of $p_k$:
\be
\Big<S_P\{\hat W^{(-)}_k\}\cdot S_R\{p_k\}\Big>^{\bf A}=\int d\mu^{\bf A}S_P\{\hat W^{(-)}_k\}\cdot S_R\{p_k\}=\int d\mu^{\bf A}S_P\{{\overleftarrow{W^{(0)}}}\}\cdot S_R\{p_k\}=\int d\mu^{\bf A}K_P\{p_k\}\cdot S_R\{p_k\}
\ee
where we use the left arrow to denote the operator acting to the left.
Thus, one could construct this way the map $S_P\{{\overleftarrow{W^{(0)}}}\}\longrightarrow K_P\{p_k\}$.
However, there is a problem in doing this.

To see the problem, let us consider the simplest case of the operator $\hat W_{[1]}^{(0)}$, which is manifestly given by
\be
\hat W_{[1]}^{(0)}=\sum_{a=1}^\infty ap_a{\p\over\p p_a}
\ee
When integrating by parts, this operator acts to the measure (\ref{Gm}) as
\be
\overleftarrow{W_{[1]}^{(0)}}=-\sum_{a=1}^\infty a{\p\over\p p_a}\cdot p_a=-\sum_{a=1}^\infty ap_a{\p\over\p p_a}-\sum_{a=1}^\infty a
\ee
The last sum requires some regularization $\sum_{a=1}^\infty a\longrightarrow\sigma_{reg}$, and, when acting to the measure, this operator produces a function, not a polynomial. Still, the Schur functions are eigenfunctions of this operator. If one acts to the Gaussian measure (\ref{Gm}) with this operator, one obtains
\be
\left(-\sum_{a=1}^\infty ap_a{\p\over\p p_a}-\sigma_{reg}\right)\cdot\underbrace{\prod_{k=1}^\infty
dp_k\exp\left(-\frac{(p_{2k-1}-1)^2}{2(2k-1)} - \frac{p_{2k}^2}{4k}\right)}_{d\mu}
=\left(\sum_{k=1}^\infty (p_k^2-p_{2k-1})-\sigma_{reg}\right)\cdot d\mu
\ee
i.e.
\be
K_{[1]}\{p_k\}=\sum_{k=1}^\infty (p_k^2-p_{2k-1})-\sigma_{reg}
\ee
At the second level, the formulas are even more involved. Indeed, consider the operator $\overleftarrow{W^{(0)}_{[2]}}$, which is given by
\be
\overleftarrow{W^{(0)}_{[2]}}=\sum_{a,b}^\infty abp_{a+b}{\p^2\over\p p_a\p p_b}-\sum_{a,b}^\infty (a+b)p_ap_b{\p\over\p p_{a+b}}
\ee
In this case, there are no infinite sums emerging from differentiating within the operator, however, in contrast with the standard $\hat W^{(0)}_{[2]}$-operator \cite{GJ,MMN},
\be
\overrightarrow{W^{(0)}_{[2]}}=\sum_{a,b}^\infty abp_{a+b}{\p^2\over\p p_a\p p_b}+\sum_{a,b}^\infty (a+b)p_ap_b{\p\over\p p_{a+b}}
\ee
whose eigenfunctions are the Schur functions, the eigenfunctions of $\overleftarrow{W^{(0)}_{[2]}}$ do not form a complete basis in the space of graded polynomials of $p_k$. For instance, at grading 3,
\be
\overleftarrow{W^{(0)}_{[2]}}S_R=A_{RP}S_P
\ee
where the matrix
\be
A_{RP}=\left(
\begin{array}{ccc}
0&-4&2\cr
4&0&-4\cr
-2&4&0
\end{array}
\right)
\ee
is degenerate: it has rank 2. Here we labelled the indices: $[3]\to 1$, $[21]\to 2$, $[111]\to 3$.

The action of $\overleftarrow{W^{(0)}_{[2]}}$ to the Gaussian measure (\ref{Gm}) is now cubic in $p_k$:
\be
\left(\sum_{a,b}^\infty abp_{a+b}{\p^2\over\p p_a\p p_b}-\sum_{a,b}^\infty (a+b)p_ap_b{\p\over\p p_{a+b}}\right)d\mu
=\\
=\left(4\sum_{k=1}^\infty p_k^2p_{2k}-4\Big(\sum_{k=1}^\infty p_{2k-1}\Big)\Big(\sum_{k=1}^\infty p_{2k}\Big)
+2\sum_{a>b\ge 1}^\infty p_ap_bp_{a+b}\right)\cdot d\mu
\ee
i.e.
\be
K_{[2]}=4\sum_{k=1}^\infty p_k^2p_{2k}-4\Big(\sum_{k=1}^\infty p_{2k-1}\Big)\Big(\sum_{k=1}^\infty p_{2k}\Big)
+2\sum_{a>b\ge 1}^\infty p_ap_bp_{a+b}
\ee
This is a general rule: any $K_R$ is a polynomial of degree $|R|+1$ in $p_k$, however, it involves $p_k$ with arbitrary large index $k$, and also involves infinite sums, which require a regularization. All this requires a careful treatment, which we postpone for a separate publication.

\section{Relation of sum formulas to Ramanujan sums}

Another amusing fact about the two nice cases (\ref{0powerdR}) and (\ref{1powerdR})
is their possible relation to the celebrated Ramanujan sums  \cite{Ram}
\be
c_{m,k}:= \sum_{(a,m)=1}^k e^{\frac{2\pi i ak}{m}}
\ee
where the sum goes over $a$, which are coprime with $m$,
and it is always a positive integer.
Counterparts of (\ref{0powerdR}) and (\ref{1powerdR}),
not literal, still surprisingly similar
appear in two ways: with plethystic and ordinary exponentials.

Defining the plethystic exponential of a function $f(x)$ as
\be
P.E.\Big(f(q)\Big):=\exp\left(\sum_{k=1}^\infty{f(q^k)\over k}\right)
\ee
one obtains
\be
 P.E.\left({q\over 1-q}\right) = \xi_0(q) \nn\\
 P.E.\left(\sum_{k=1}^\infty {c_{2,k}q^k\over k}\right) = e^{q^2+q} = \sqrt{\xi_1(2q)}   \nn\\
P.E.\left(\sum_{k=1}^\infty {c_{1,k}q^k\over k}\right) = e^q = \xi_2(q)
\ee
where
\be
\xi_m(q):=\sum_Rq^{|R|}d_R^m
\ee
is the quantity (\ref{dimrep}).
In terms of ordinary exponentials, one gets
\be
\xi_0(q)=\exp\left(\sum_{k=1}^\infty{q^k\over k(1-q^k)}\right)\nn\\
\xi_1(q)=\exp\left(\sum_{k=1}^\infty{c_{2,k}q^k\over 1-q^k}\right)=e^{2q^2+q}\nn\\
\xi_2(q)=\exp\left(\sum_{k=1}^\infty{c_{1,k}q^k\over 1-q^k}\right)=e^q
\ee

In fact,
\be
P.E.\left(\sum_{k=1}^\infty {c_{0,k}q^k\over k}\right) =\exp\left({q\over 1-q}\right)
\nn \\
\exp\left(\sum_{k=1}^\infty{c_{0,k}q^k\over 1-q^k}\right)=\exp\left({q\over (1-q)^2}\right)
\ee
and
\be
P.E.\left(\sum_{k=1}^\infty {c_{m,k}q^k\over k}\right) = \exp\left(\sum_{d|m} q^d\right),\ \ \ \ \ m>0 \nn \\
\exp\left(\sum_{k=1}^\infty{c_{m,k}q^k\over 1-q^k}\right)= \exp\left(\sum_{d|m} d\cdot q^d\right),\ \ \ \ \ m>0
\ee
are always expressed through the exponentiated finite sum over divisors of $m$.
Thus, these are simple formulas, which, unfortunately, have nothing to do with our $\xi_m$
at $m>2$.

\section{Hurwitz tau-functions}

The combinatorial Borel transform allows one to enrich the set of the Hurwitz $\tau$-functions \cite{MMN,AMMN}
\be
\tau^{(n)}\{p^{(1)},\ldots,p^{(n)}\}:=\sum_R q^{|R|} d_R^{2-n} S_R\{p^{(1)}\}\ldots S_R\{p^{(n)}\}
\ee
to make the power of $d_R$ and the number of Schur functions independent: suppose $m>n$, then
\be
\sum_R d_R^{2-m} S_R\{p^{(1)}\}\ldots S_R\{p^{(n)}\}
= \left<\tau^{(m)}\{p^{(1)},\ldots,p^{(n)},p^{(n+1)}, \ldots, p^{(m)}\}\right>_{p^{(n+1)}, \ldots , p^{(m)}}
\ee
Averaging here is performed over the $m-n$ sets of time variables.
For example, a direct counterpart of (\ref{factorialelimination}) would be
\be
\sum_R \frac{q^{|R|}}{d_R}
= \sum_R  \frac{q^{|R|}}{d_R}\underbrace{\left<S_R\{p^{(1)}\right>}_{1}\underbrace{\left<S_R\{p^{(2)}\right>}_{1}
\underbrace{\left<S_R\{p^{(3)}\right>}_{1}
 = \left< \tau^{(3)}\{p^{(1)},p^{(2)},p^{(3)}\}\right>_{p^{(1)},p^{(2)},p^{(3)}}
\ee
At the same time, from what we know about the spectrum of dimensions,
\be
\sum_R \frac{q^{|R|}}{d_R} = \sum_R q^{|R|}\frac{|R|! }{{\rm dim}_R} =
 \sum_{r=0}^\infty r! \cdot q^{r}\left\{2 + 2\cdot\frac{1}{r-1}
 + 2\cdot \left( \frac{1}{\frac{r(r-3)}{2}} +  \frac{1}{\frac{(r-1)(r-2)}{2}}\right) +
\right.\nn \\ \left.
+  2\cdot \left( \frac{1}{\frac{r(r-1)(r-5)}{6}} +  \frac{1}{\frac{r(r-2)(r-4)}{3}}
+ \frac{1}{\frac{(r-1)(r-2)(r-3)}{6}}\right)
  + O(r^{-4})  \right\}
\label{dinvsum}
\ee
This first item corresponds to representations $[r]$ and $[1^{r}]$,
the second, to $[r-1,1]$ and $[2,1^{r-2}]$ and so on:
\be
{\rm dim}_{[r-1,1]}=  r-1\nn \\ \nn \\
{\rm dim}_{[r-2,2]}={r(r-3)\over 2}\nn\\
{\rm dim}_{[r-2,1,1]}={(r-1)(r-2)\over 2}\nn\\ \nn \\
{\rm dim}_{[r-3,3]}={r(r-1)(r-5)\over 6}\nn\\
{\rm dim}_{[r-3,2,1]}={r(r-2)(r-4)\over 3}\nn\\
{\rm dim}_{[r-3,1,1,1]}={(r-1)(r-2)(r-3)\over 6 } \nn \\ \nn \\
\ldots
\label{listdims}
\ee
It is important that this is a discrete spectrum and multiplicities stabilize at large $r$.
The smallest dimensions, like listed in (\ref{listdims}) are made from polynomials like
$\frac{r^m}{m!}\Big(1+O(r^{-1})\Big)$,
but at another end of the spectrum the growth is already factorial.
In particular,
the biggest dimension $D_r$ of all ${\rm dim}_R$ at the given level $|R|=r$ is restricted  at large $r$ by \cite{McKay}:
\be
\sqrt{r!}e^{-\alpha_1\sqrt{r}(1+o(1))}\le D_r\le \sqrt{r!}e^{-\alpha_2\sqrt{r}(1+o(1))}
\ee
where $\alpha_{1,2}$ are some constants.
Still this growth is much weaker than $r!$ in the numerator  of (\ref{dinvsum}).
Thus contributions with all $m$ are divergent at large $r=|R|$,
and one can expect the whole series to be ambiguous.
The same is true for any particular term in the expansion (\ref{dinvsum}).

We will deal with the divergent sums $ \xi_{-\nu} =\sum_R \frac{q^{|R|}}{d_R^\nu} $ with negative powers of $d_R$ as with formal power series. Note that they have {\it integer}-valued coefficients, provided by
inverse powers of representation dimensions $d_R =\frac{|R|!}{{\rm dim}_R}$
with multiplicities read from (\ref{xim}).
Their plethystic logarithms $\Psi_\nu(k)$ also have an interesting interpretation. Indeed, consider
\be
\xi_{-\nu} =
\sum_R \frac{q^{|R|}}{d_R^\nu} =\exp\left(\sum_{k=1}^\infty{\Psi_\nu(k)q^k\over k}\right)
\ee
Remarkably, all the coefficients $\Psi_\nu(k)$ are positive integers,
and they can be interpreted as the numbers of subgroups of index $k$
in the fundamental group of a certain fiber space \cite{LM}.
They are a little more complicated (multi-linear) combinations of powers of above-mentioned
integers $d_R =\frac{|R|!}{{\rm dim}_R}$ so that in the decomposition
\be\label{Psi}
\Psi_\nu(k) = \sum_{n=1}^\infty c_n(k)\eta_n^\nu
\ee
the spectrum $\eta_n$ of $\Psi_\nu(k)$ is more complicated than the spectrum $d_R^{-1}$ of $\xi_{-\nu}$, with some of elements $\eta_n$ coming from products of $d_R^{-1}$, and the coefficients $ c_n(k)$ are not obligatory positive (despite $\Psi_\nu(k)$ {\it is} positive), and, hence, they can not play a role of multiplicities.

One can compare the spectrum of $\xi_{-\nu}$ (the values of $d_R^{-1}$),

\be
\begin{array}{c||c|c|cc|ccc|cccc|ccccc|cccc}
|R| & 1 & 2 & 3 && 4 &&& 5 &&&& 6 &&&&&   \ldots
\\ \hline
d_R^{-1} &1 &2    & 3 & 6   & 8& 12 & 24 & 20 &24 & 30&  120
& 45 & 72 & 80 & 144 & 720 &
\end{array}
\label{primaryspectrum}
\ee
with the spectrum of $\Psi_\nu(k)$ appearing in (\ref{Psi}) (see Table 3 of \cite{LM}, note that $\Psi_\nu(k)$ are denoted as $R_\nu(k)$ in \cite{LM}):

\bigskip

{\footnotesize
\be
\hspace{-1cm}\begin{array}{c||ccccccccccccccccccccccc}
\Psi_\nu(k)\Big({\rm Table}\ 3\Big) &\eta_1&\eta_2&\eta_3&\eta_4&\eta_5&\eta_6&\ldots&&&&&&&&&&&&
\\ \hline
\Psi_\nu(1) & 1 &&&&&&&&&&&&&&&&&
\\ \hline
\Psi_\nu(2) & 1&2 &&&&&&&&&&&&&&&&&&
\\ \hline
\Psi_\nu(3) & 1&2&3&&6&&&&&&&&&&&&&&&&
\\ \hline
\Psi_\nu(4) & 1&2&3&\boxed{4}&6&8&&12&24&&&&&&&&&&&
\\ \hline
\Psi_\nu(5) & 1&2&3&\boxed{4}&&8&&12&&&&20&&30&&&&&&120&
\\ \hline
 \Psi_\nu(6)& 1&2&3&\boxed{4}&6&8&\boxed{9}&12&&\boxed{16}&\boxed{18}&20&24&30&\boxed{36}&45&\boxed{48}&72&80&120&144&720
\\ \hline
\ldots
\end{array}
\nn
\ee
}

\noindent
where the quantities in the boxes are the products of the powers of the primary elements of the spectrum
from (\ref{primaryspectrum}), and for each $\Psi_\nu(k)$ only $\eta_n$ corresponding to non-zero $c_n$ in (\ref{Psi}) are specified.

\section{Conclusion}

In this paper we raised the trick of \cite{2212.04859} to the level of a {\it method},
applicable to a large variety of functions and providing a unified treatment
of the so far sporadic set of amusing identities.
We emphasize its conceptual relation to the Borel transform, which is used in physics
to explain and put under control the ambiguity of divergent series.
As we demonstrate, the abilities of our suggested {\it combinatorial Borel transform} are much wider
and applications are not restricted to divergent series.
There are plenty of different directions to develop these ideas,
and also a number of conceptual questions to resolve, of which we mention just two.

\bigskip

{\bf Basic open questions:}

\begin{itemize}
\item[1)] When (for what conditions $\left<S_R\right> = \alpha_R$) does a simple averaging procedure exist?
When is the measure Gaussian?

\item[2)] How does a trivial Gaussian measure like (\ref{Gm}) converts non-trivial series
like $F\{p\}=\sum_R q^{|R|}d_R^2 S_R\{p\}$ into a triviality: $\left<F\{p\}\right> = e^q$?
What is the role of triangularity (only $p_k$ with $k\leq |R|$ affect the coefficient of $q^{|R|}$)?
To what extent and in what sense can the averaging procedure (the combinatorial Borel transform) be invertible?

\bigskip

For example, after averaging of
{\footnotesize
\be
\sum_R q^{|R|}d_R^2S_R\{p\} = 1 +qp_1 + \frac{q^2p_1^2}{4} + \frac{q^3(5p_1^3-2p_3)}{4\cdot 27} + \frac{q^4(4p_1^4-\frac{3}{2}p_2^2-p_3p_1)}{2^5\cdot 3^3}
+ \frac{q^5(149p_1^5-80p_3p_1^2-75p_2^2p_1+36p_5)}{2^7\cdot 3^3\cdot 5^3} + \ldots
\nn
\ee
}
with measure {\bf A} (\ref{Gm}) over all $p_k$ but $p_1$, the last step of averaging is
{\footnotesize
\be
e^q = \int_{-\infty}^\infty \frac{e^{-\frac{(p_1-1)^2}{2}} dp_1}{\sqrt{2\pi}}
\left(1 +qp_1 + \frac{q^2p_1^2}{4} + \frac{q^3(5p_1^3-2)}{4\cdot 27} + \frac{q^4(4p_1^4-p_1-3)}{2^5\cdot 3^3}
+ \frac{q^5(149p_1^5-80p_1^2-150p_1+36)}{2^7\cdot 3^3\cdot 5^3}
+\ldots \right)
\nn
\ee
}

\noindent
Clearly this integration is already radical enough to allow for any simple inversion.

\item[3)] In the paper, we only started exploration of the combinatorial Borel transform. In particular, we did not discuss the role of analytic continuation within this context. Moreover, we did not discuss in detail when the sums under consideration have different possible types of growth, in particular, a factorial divergence and other types of divergencies (for exception of comments in sec.6). This is especially important in the case of the Hurwitz $\tau$-functions in order to reveal the non-perturbative nature of the large order behaviour in these.
\end{itemize}

\bigskip

We hope that this new, more general look on a combinatorial generalization of the Borel transform would lead to new insights or, at least, to new amusing relations and formulas, which look artificial from other points of view. There are many of this kind, besides the ones that we mentioned, and now we have a new option to interpret and understand at least some of them.

\section*{Acknowledgements}

We are grateful to the referee of our paper for valuable comments. This work was supported by the Russian Science Foundation (Grant No.23-41-00049).

\end{document}